# A HUBBLE SPACE TELESCOPE CATALOG OF 449 GALAXIES SEEN THROUGH THE DISK OF M31

Paul Hodge
(Astronomy Department, University of Washington, Seattle, WA, 98195, USA, hodge@astro.washington.edu)

and

Karl Krienke (Seattle Pacific University, Seattle, WA98119, okk@spu.edu)

## ABSTRACT

From inspection of 30 Hubble Space Telescope ACS images of M31, we provide a catalog of 449 galaxies seen through the spiral disk. Measurements of the positions of the galaxies, their integrated magnitudes in two colors and their sizes, determined from isophotometry, are included in the catalog. We discuss the many difficulties of interpreting these data in terms of the effects of intervening extinction by dust in the disk.

## 1. INTRODUCTION

Ever since the era in which it was recognized that galaxies contain dust that causes extinction of light, it has been tempting to measure this dust by examination of more distant galaxies seen in the background. The first published account of this technique that we are aware of was made by Shapley and Nail (1951), who used galaxy counts made in the direction of the Small Magellanic Cloud to conclude that it is essentially transparent. As an intimation of this technique's troubled future, their result was wrong, apparently because of an unfortunate accident. As related by Hodge (1974), more recent evidence showed that Shapely and Nail used number counts of non-stellar objects detected by their assistants on plates taken by the Boyden Observatory's 1.5 m telescope. Apparently they didn't examine these plates themselves, as a repeat of this experiment showed that inside the boundaries of the SMC, most of the non-stellar objects marked in ink on the plates were either emission nebulae or star clusters. The number of such objects approximately equaled the number of galaxies not seen behind the SMC. The story does not end there, however. In recent experiments, cited below, it has been shown that the crowding of stellar images in a galaxy's disk can obscure background galaxies to a considerable extent. (Incidentally, Hubble (1934) worried about this problem in his study of the Milky Way's dust in the Zone of Avoidance). It is therefore likely that Shapley and Nail's result was approximately correct, after all, because of two opposing mistakes.

More recent attempts to use this technique have been more successful. Among others, the pioneering work of Gonzalez et al. (1998), which has continued in a series of important papers (see Holwerda et al. 2007), has demonstrated that information on the opacity of galaxy disks can be obtained from background galaxy counts, when careful measurements of the various systematic effects are carried out. They say, however, that the technique should not be used for galaxies in the Local Group, where the resolved stellar foreground interferes. Somewhat contrary results were published by us (Krienke and Hodge 2001), who showed that rather weak results could be obtained for three Local Group galaxies, but, among other things, the distances to the individual background galaxies were required to be able to make reliable K corrections to their colors. Another example is the work of Dutra et a. (2001), who found that the use of background galaxy colors combined with redshifts gave some useful values for the reddening in parts of the Large and Small Magellanic Clouds.

One can argue that it does not seem to make sense that a procedure that gives information for distant galaxies does not work for nearby galaxies, where the total obtainable information is orders of magnitude greater. This is not an uncommon problem in astronomy, where we are always near the limit of the impossible. For distant objects the difficulties are smoothed out, permitting the anomalous signals to be recognized and measured more easily. For the case of the highly-resolved Local Group galaxies, the same data are available, but their completeness makes the task of extracting anything from it more complicated and the uncertainties involved are more obvious and thus more daunting.

One can conceive of using three different sets of data to detect the effects of extinction by dust in a nearby galaxy's disk using background galaxies:

a. An excess reddening of the colors,
b. Decreased integrated magnitudes, and
c. A decreased number of galaxies detected

An even larger list can be developed for the difficulties involved in using these data:

d. to measure an excess color, one must know the intrinsic color, which requires knowing the Hubble type of the galaxy and keeping in mind the intrinsic spread in the type-color relation.
e. Also one must know the distance and/or radial velocity of the galaxy in order to apply the K corrections due to the redshift of the galaxies' SEDs. This must be known also to correct the effective characteristics of the galaxies because of relativistic fading at larger values of z.
f. to detect a decrease in the magnitudes of background galaxies, one must know the distribution of magnitudes of galaxies external to M31 and its dispersion, which is large, even not including effects of clustering
g. to use the number of galaxies per area behind M31, one must establish a uniformity of sample, corrected for detection limits and distance (including relativistic effects)
h. All of the methods are complicated by the presence of a variable and often dense field star population, which affects the detection limits as well as the measurements, in a number of ways.

Because of these many difficulties, the problem is daunting one, but not completely impossible. We report here a possible method of measuring the extinction of background galaxy light by the disk of M31, one of the worst environments in the Local Group in which to use this technique. We were tempted to do this by noting the wealth of galaxies detected in the M31 fields with which we worked on M31's open clusters (Krienke and Hodge 2007, 2008). Perhaps, with the availability of hundreds of background galaxies, it might be worthwhile to make this attempt. As we point out below, the problem probably can be solved, but only with additional data, especially radial velocity data.

Please note that this paper deals only with the galaxy count problem; alternate measurements of the dust content of M31 are coming from infrared Spitzer data (Gordon et al., 2004)

## 2. OBSERVATIONAL MATERIAL

From the Hubble Space Telescope Archives, we have chosen 28 ACS pointings that cover a variety of positions in the main disk of M31, for which the archives include at least two colors that can be transformed into standard responses, Johnson-Cousins B, V and I. In addition we chose two pointings that are near but not in the main disk of the galaxy. The pointings and their positions are given in Krienke and Hodge (2008). We chose ACS data and not WFPC2 data because we felt that the increased resolution and depth of ACS would be important.

We chose to select galaxies by eye. This is defensible in the case of M31, which has densely-packed field stars, nebulae and star clusters on most of the selected pointings. We argue that using one of the available automatic selection extractors would produce so many false candidates that we would have to pick them over by eye in any case, re-introducing any selection effect that was there before the extractor process.

## 3. GALAXY PROPERTIES

Tables 1 and 2 (at end of this paper) provide catalogs of 449 galaxies found on the 30 pointings examined. We give measures of the magnitudes and colors (B, V, and I, transformed from the HST filters), their uncertainties, the J2000 positions of the centers of the galaxy images and an approximate size parameter, which is the maximum value of the distance between the center and the largest distance from the center found on a limiting isophote. The galaxy photometry was carried out with a program that one of us wrote in IDL that makes special allowance for variable "sky" values caused by the variable stellar density in the field. It was adapted from that described in our work on open clusters (Krienke and Hodge 2007).

Galaxy magnitudes ranged from V = 16 to 25, with the majority fainter than V = 19. Figures 1a and 1b show color-magnitude diagrams for the set with V and I colors and with B and V colors, respectively. The distribution of points in these diagrams resembles that found for galaxy surveys in general.

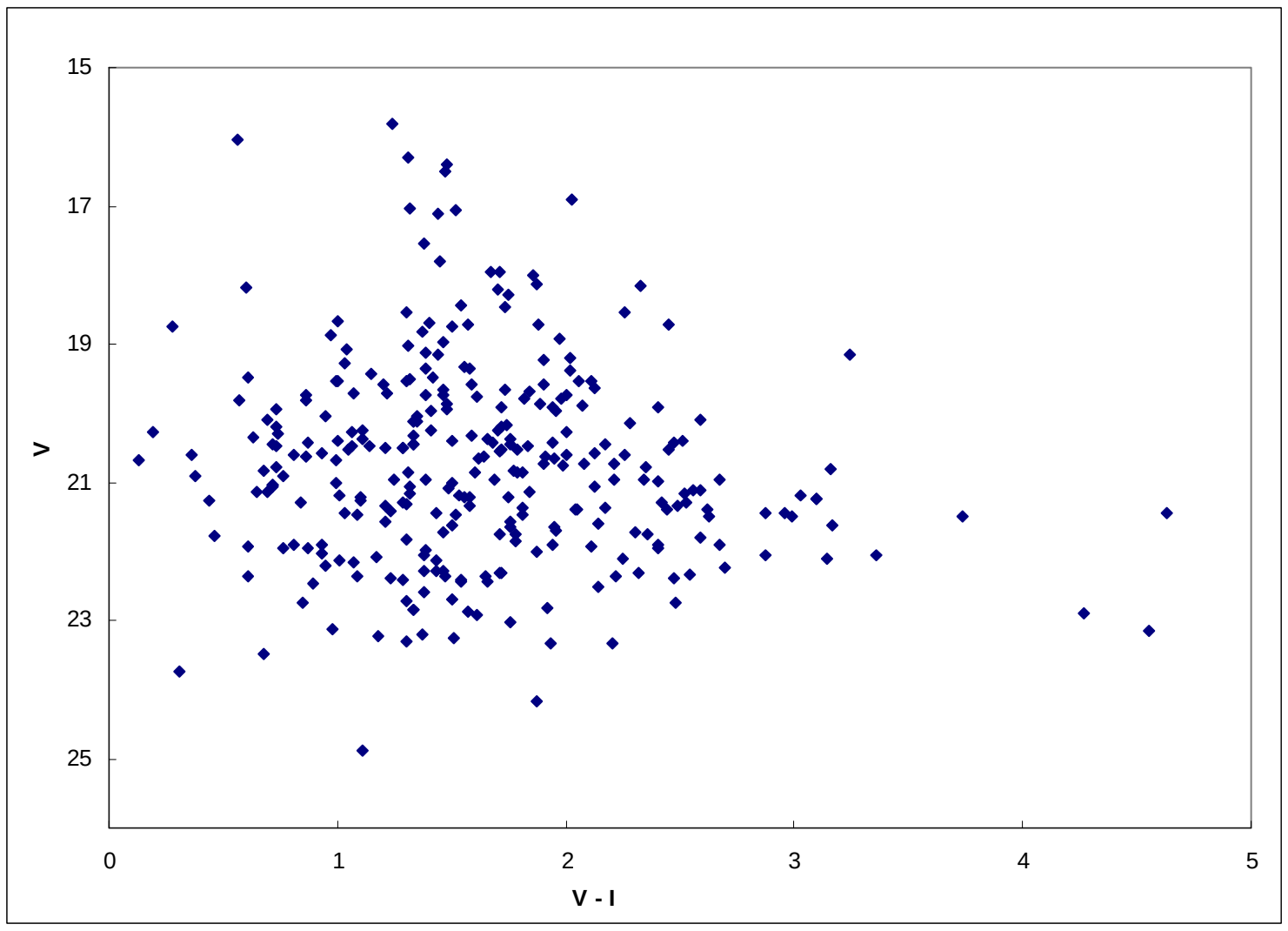

Fig. 1a. The V, I color-magnitude diagram

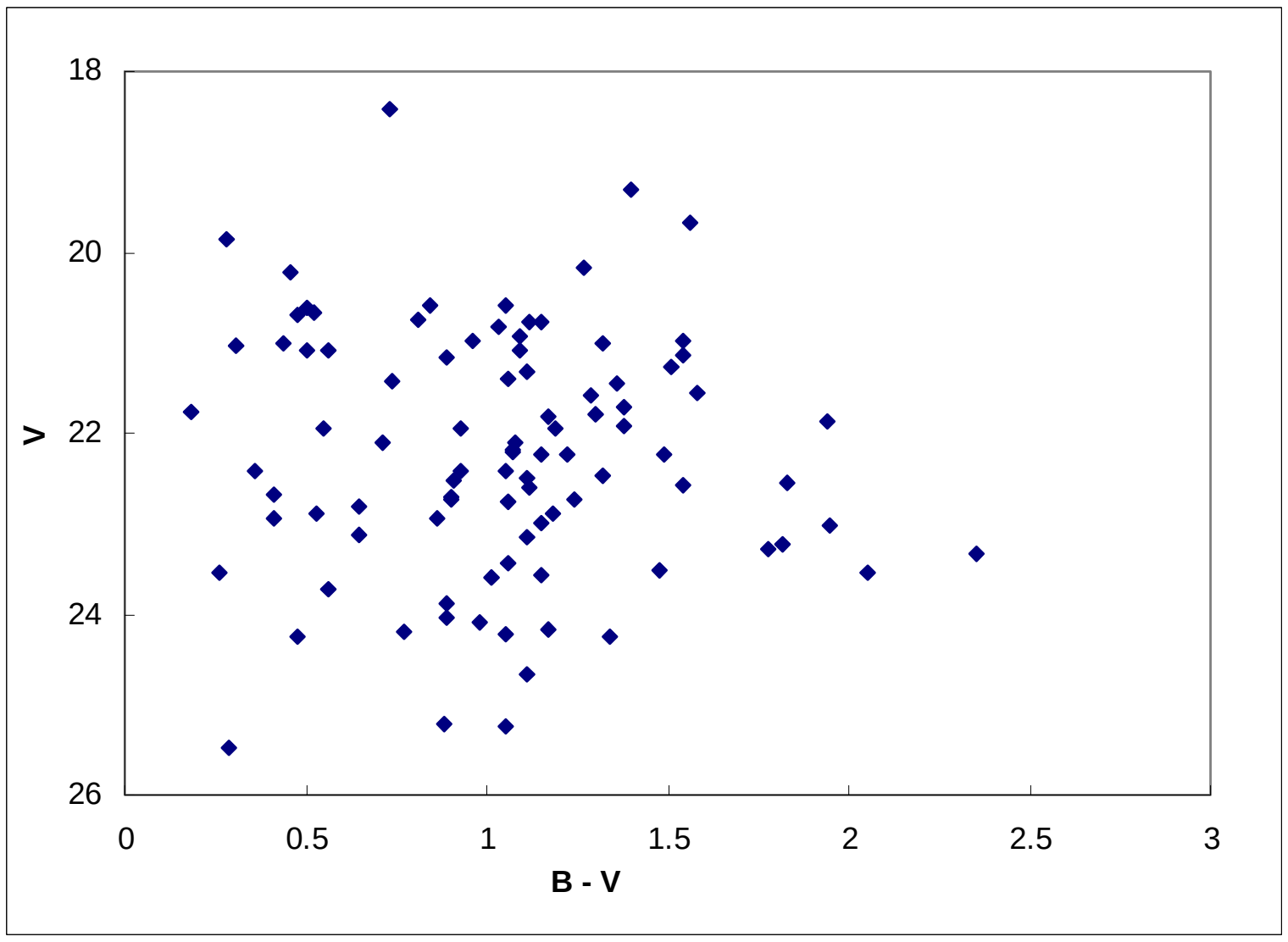

Fig. 1b. The B, V color-magnitude diagram.

The quoted errors are primarily dominated by the variable background. Figure 2 shows the distribution of errors in the V measurements. Average values of the errors are B(error) = 0.18 ,V(error) = 0.14, I(error) = 0.11.

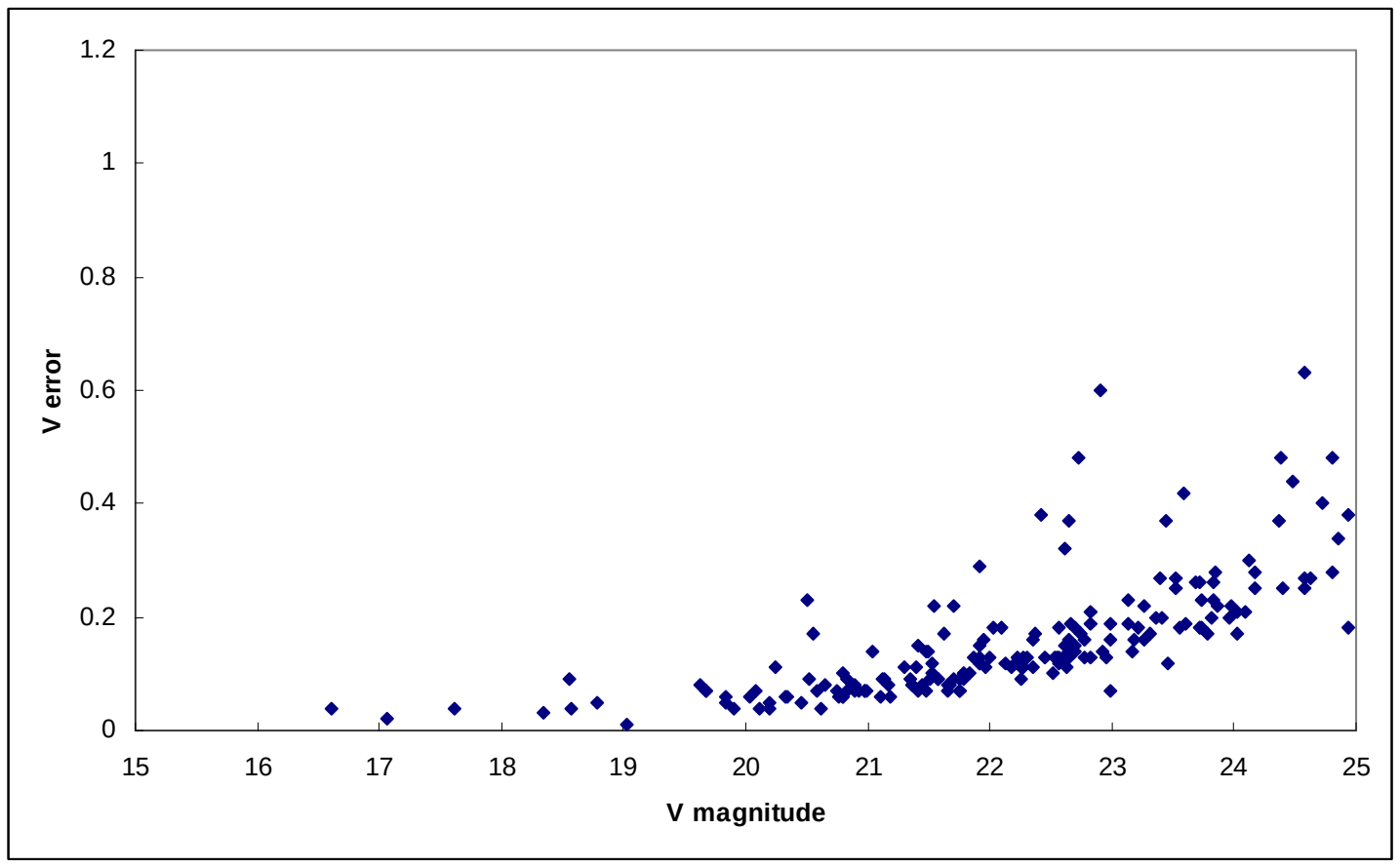

Fig. 2 Measurement errors for V magnitudes

We also scanned three ACS pointings that lie outside of but near the M31 disk. One of these includes a distant cluster of galaxies and the other two were obtained to study two outer globular clusters. Figure 3 shows a portion of one the outer fields, including several galaxies plus the M31 globular cluster Bol 409.

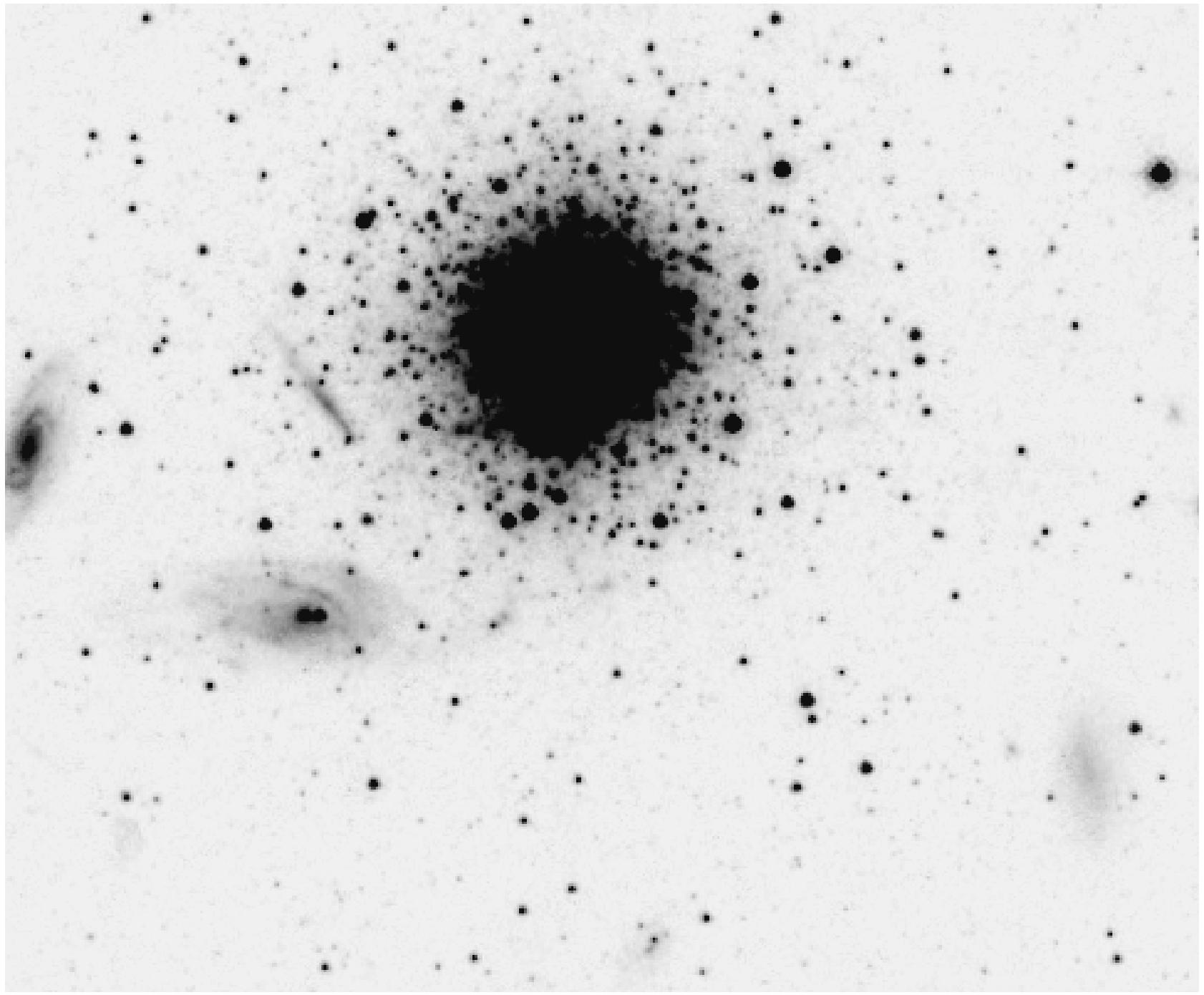

Fig. 3. This image from one of the outer pointings shows four galaxies, plus the globular cluster Bol409 at J2000 = 12.541258, 41.683533. The globular has a V magnitude of 16.09 and a V – I of 1.24.

In Figure 4 we have plotted the distribution in color of the galaxies in our M31 sample, for the full sample and for the bright and faint galaxies separately. From this figure it is clear that the color distribution of galaxies is relatively insensitive to changes in magnitude, except for an excess of faint galaxies among the redder colors.

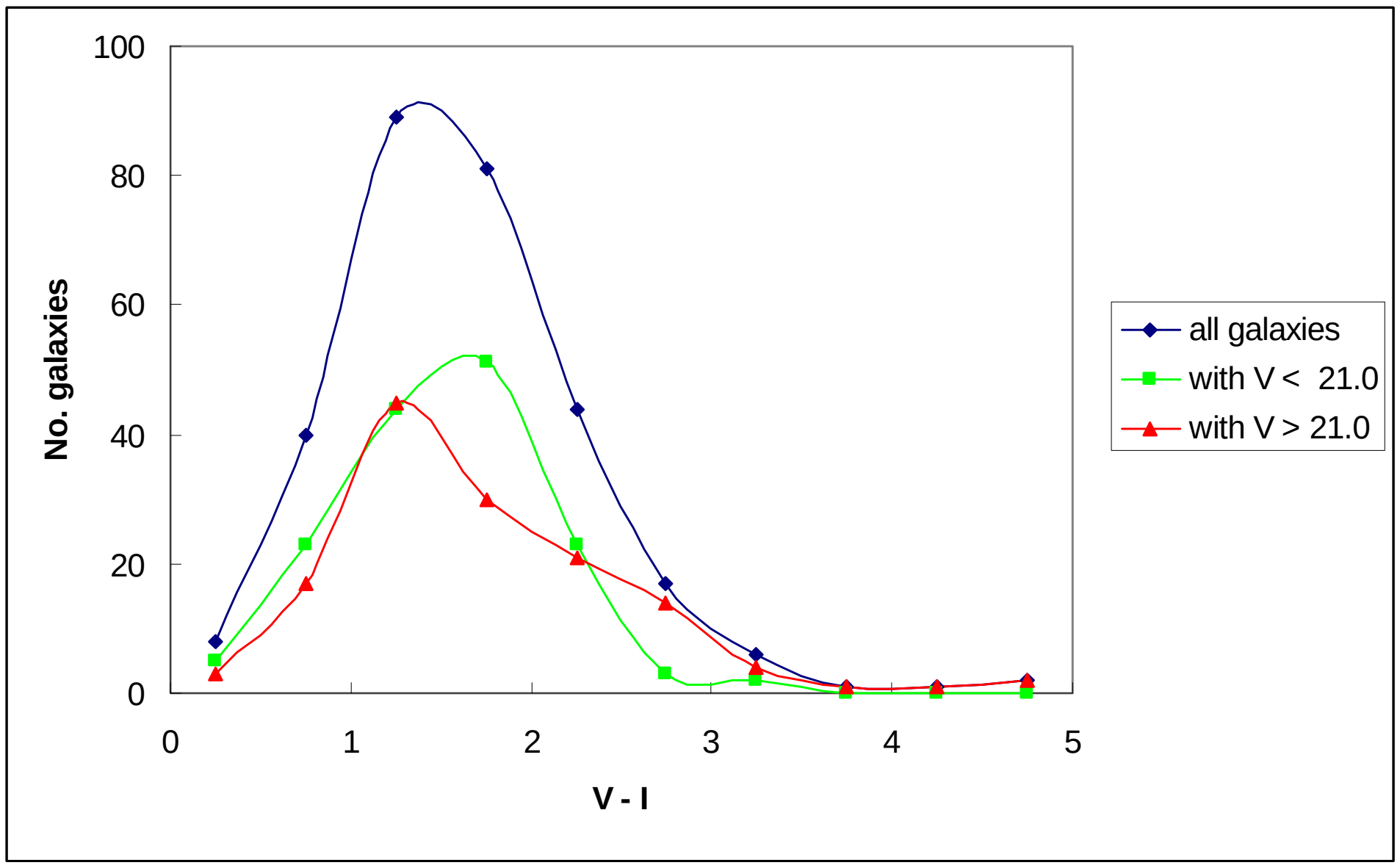

Figure 4. The distribution of galaxy colors. The fainter galaxies show an excess of red colors.

Two causes can be considered to explain this effect: (1) statistically the fainter clusters will have greater red-shifts and thus redder colors and (2) galaxies reddened by M31's interstellar dust will tend to be fainter because of absorption. Without a measure of each galaxy's redshift, it is very difficult to determine how much of the reddening is due to each of these effects.

## 4. CLASS A GALAXIES

We have selected 42 galaxies that are bright enough and well enough resolved to determine approximate Hubble types for them. These are listed in Table 3, together with their properties, which were determined from repeated measurements, made after the main photometric program (Tables 1 and 2) was competed. These galaxies are the best samples for eventual analysis, as knowing their Hubble types narrows down the possible values of their intrinsic properties.

TABLE 3. CLASS A GALAXIES

| No. | V | V err | I | I err | RA | Dec | a arcsec | class |
|---|---|---|---|---|---|---|---|---|
| 1 | 20.93 | 0.06 | 19.36 | 0.05 | 9.88307 | 40.76366 | 0.8 | Sa |
| 2 | 17.92 | 0.03 | 16.44 | 0.02 | 9.89667 | 40.80440 | 0.5 | Sc |
| 3 | 20.24 | 0.06 | 18.95 | 0.05 | 10.46134 | 41.41127 | 0.4 | Scp |
| 4 | 17.96 | 0.06 | 16.51 | 0.04 | 10.46558 | 41.41164 | 0.3 | Sc |
| 5 | 21.23 | 0.1 | 19.69 | 0.04 | 10.47878 | 41.44724 | 0.4 | Sc/Irr |
| 6 | 19.46 | 0.07 | 17.76 | 0.05 | 10.48474 | 41.42366 | 0.2 | Sa |
| 7 | 17.80 | 0.07 | 16.46 | 0.02 | 10.50661 | 41.41751 | 0.5 | SBc |
| 8 | 18.84 | 0.02 | 17.32 | 0.02 | 10.52896 | 41.05360 | 0.5 | Sbc |
| 9 | 22.49 | 0.09 | 20.76 | 0.13 | 10.53237 | 41.01934 | 0.1 | S0 |
| 10 | 20.44 | 0.04 | 19.01 | 0.06 | 10.54467 | 41.05218 | 0.2 | Sc |
| 11 | 21.30 | 0.12 | 19.43 | 0.07 | 10.55152 | 41.50057 | 0.2 | Sbc |
| 12 | 21.67 | 0.12 | 19.82 | 0.09 | 10.55680 | 40.92756 | 0.4 | E |
| 13 | 20.98 | 0.1 | 19.56 | 0.1 | 10.56445 | 40.95871 | 0.3 | Sc |
| 14 | 24.66 | 0.52 | 22.72 | 0.25 | 10.56690 | 41.39300 | 0.4 | Sc? |
| 15 | 22.46 | 0.1 | 20.45 | 0.08 | 10.57412 | 41.55530 | 0.1 | Sc |
| 16 | 20.52 | 0.07 | 18.59 | 0.03 | 10.58587 | 40.95334 | 0.9 | Sc |
| 17 | 20.70 | 0.07 | 19.49 | 0.05 | 10.58954 | 41.50013 | 0.1 | Sc |
| 18 | 20.01 | 0.06 | 18.15 | 0.03 | 10.59522 | 40.95309 | 0.2 | Sc |
| 19 | 21.22 | 0.09 | 19.42 | 0.06 | 10.63844 | 41.59062 | 0.0 | E |
| 20 | 24.55 | 0.31 | 21.79 | 0.13 | 10.66056 | 41.06394 | 0.2 | E2 |
| 21 | 18.45 | 0.04 | 16.90 | 0.04 | 10.66994 | 41.47275 | 0.2 | SBc |
| 22 | 21.09 | 0.13 | 19.43 | 0.05 | 10.67934 | 41.60216 | 0.1 | Sb |
| 23 | 21.24 | 0.09 | 19.79 | 0.12 | 10.68106 | 41.59905 | 0.2 | Sc |
| 24 | 23.06 | 0.21 | 21.15 | 0.18 | 10.68244 | 41.60091 | 0.2 | Sb |
| 25 | 19.63 | 0.07 | 18.31 | 0.04 | 10.72812 | 41.69887 | 0.1 | Sc |
| 26 | 22.31 | 0.15 | 20.46 | 0.09 | 10.72963 | 41.52676 | 0.1 | E0 |
| 27 | 20.83 | 0.13 | 18.89 | 0.05 | 10.73232 | 41.68054 | 0.1 | E3 |
| 28 | 20.54 | 0.09 | 19.62 | 0.05 | 10.73407 | 41.15078 | 0.3 | Sb? |
| 29 | 25.89 | 0.83 | 24.45 | 0.4 | 10.74074 | 40.96886 | 0.2 | Sc? |
| 30 | 21.09 | 0.12 | 19.77 | 0.05 | 10.74235 | 41.70736 | 0.2 | Sc/Irr |
| 31 | 22.60 | 0.15 | 21.45 | 0.15 | 10.75560 | 41.70253 | 0.2 | E5 |
| 32 | 22.07 | 0.24 | 21.28 | 0.19 | 10.75873 | 40.95387 | 0.0 | E2 |
| 33 | 24.62 | 0.38 | 23.58 | 0.48 | 10.75873 | 40.95387 | 0.1 | ? |
| 34 | 19.06 | 0.12 | 18.71 | 0.15 | 10.76691 | 40.96826 | 0.0 | Irr |
| 35 | 18.94 | 0.1 | 16.91 | 0.04 | 10.82526 | 41.67233 | 0.0 | Sc |
| 36 | 18.92 | 0.04 | 17.41 | 0.04 | 10.84596 | 41.04025 | 0.1 | Sb |
| 37 | 19.65 | 0.06 | 18.30 | 0.07 | 10.86144 | 41.06189 | 0.2 | Sc |
| 38 | 21.11 | 0.09 | 19.18 | 0.05 | 10.88520 | 41.07125 | 0.1 | Sb |
| 39 | 19.28 | 0.06 | 17.39 | 0.04 | 10.92486 | 41.73431 | 0.0 | Scp |
| 40 | 19.63 | 0.06 | 17.96 | 0.03 | 10.95549 | 41.18157 | 0.0 | Sb |
| 41 | 19.48 | 0.04 | 17.62 | 0.03 | 10.97143 | 41.74332 | 0.1 | Sc |
| 42 | 21.73 | 0.1 | 19.70 | 0.06 | 10.98047 | 41.73055 | 0.0 | Sc |

Figure 4 shows an apparent group of Class A galaxies seen through a relatively low density section of the M31 near the minor axis on the western, farther and dustier side of M31 (Figure 5). The galaxies are well-resolved and even show fainter outer structure including an apparent tidal tail.

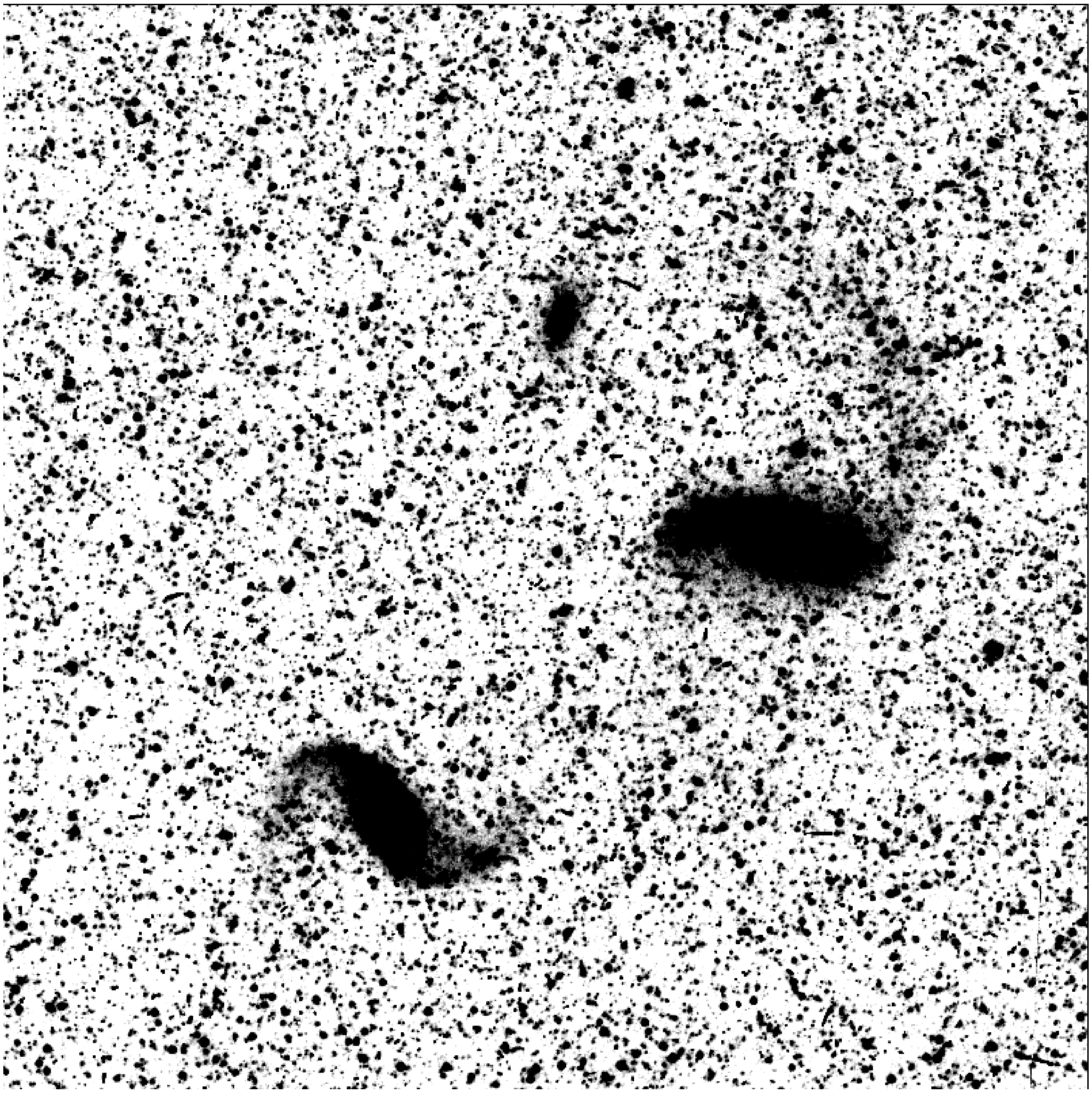

Figure 4. The Trio, an apparent group of three galaxies seen through the outer arms of M31. The galaxies are (from left to right) Nos. 109, 108 and 110. This is a V image printed to show the outer parts of the galaxies, including a tidal arm

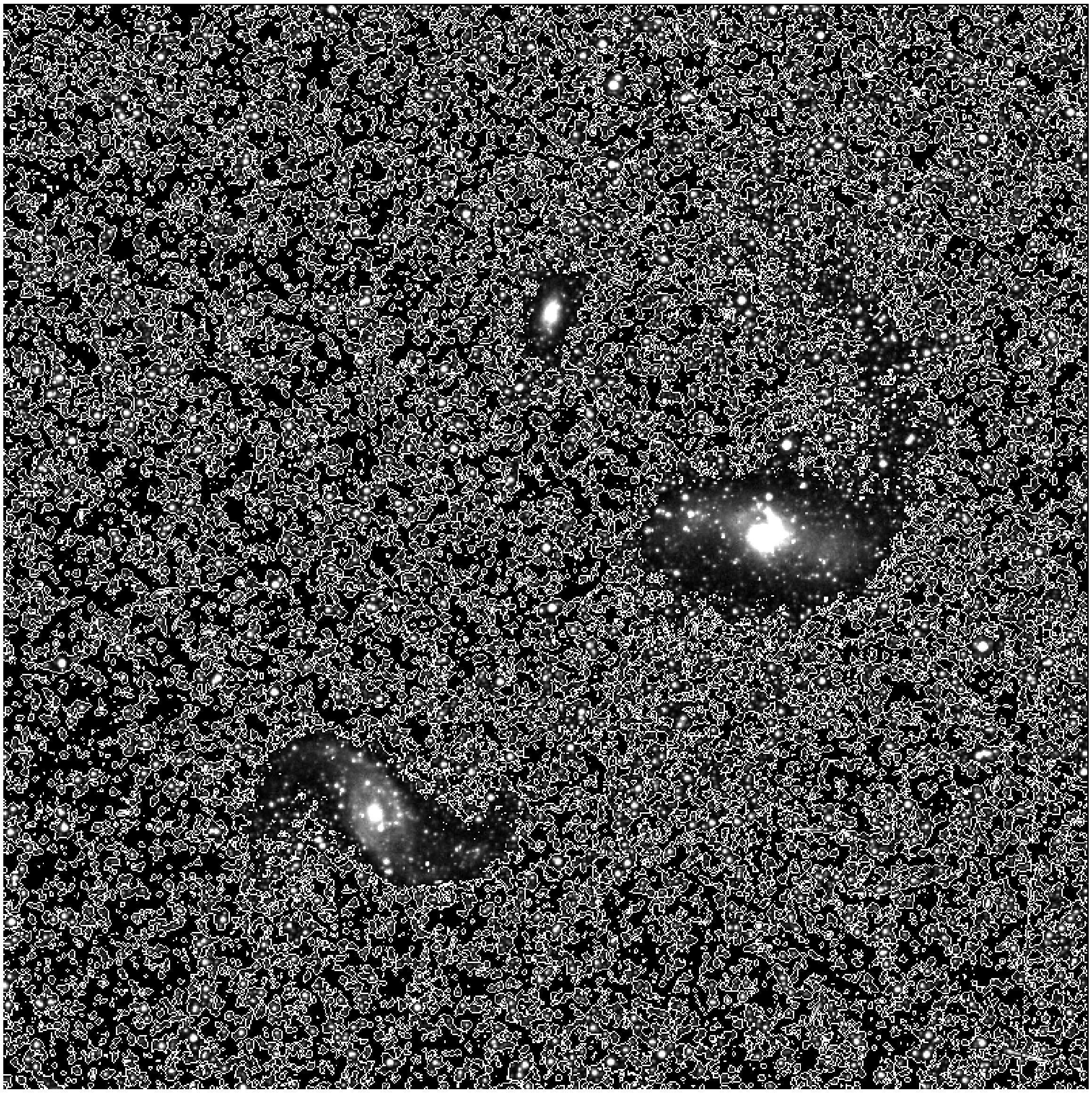

Figure 5. An isophotal map of the Trio that shows the outer probable tidal arm from Galaxy 110 towards Galaxy 108. This figure also shows the inner structure of the galaxies better than in Figure 4. The highly mottled background characterizes the rich field of M31’s disk stars.

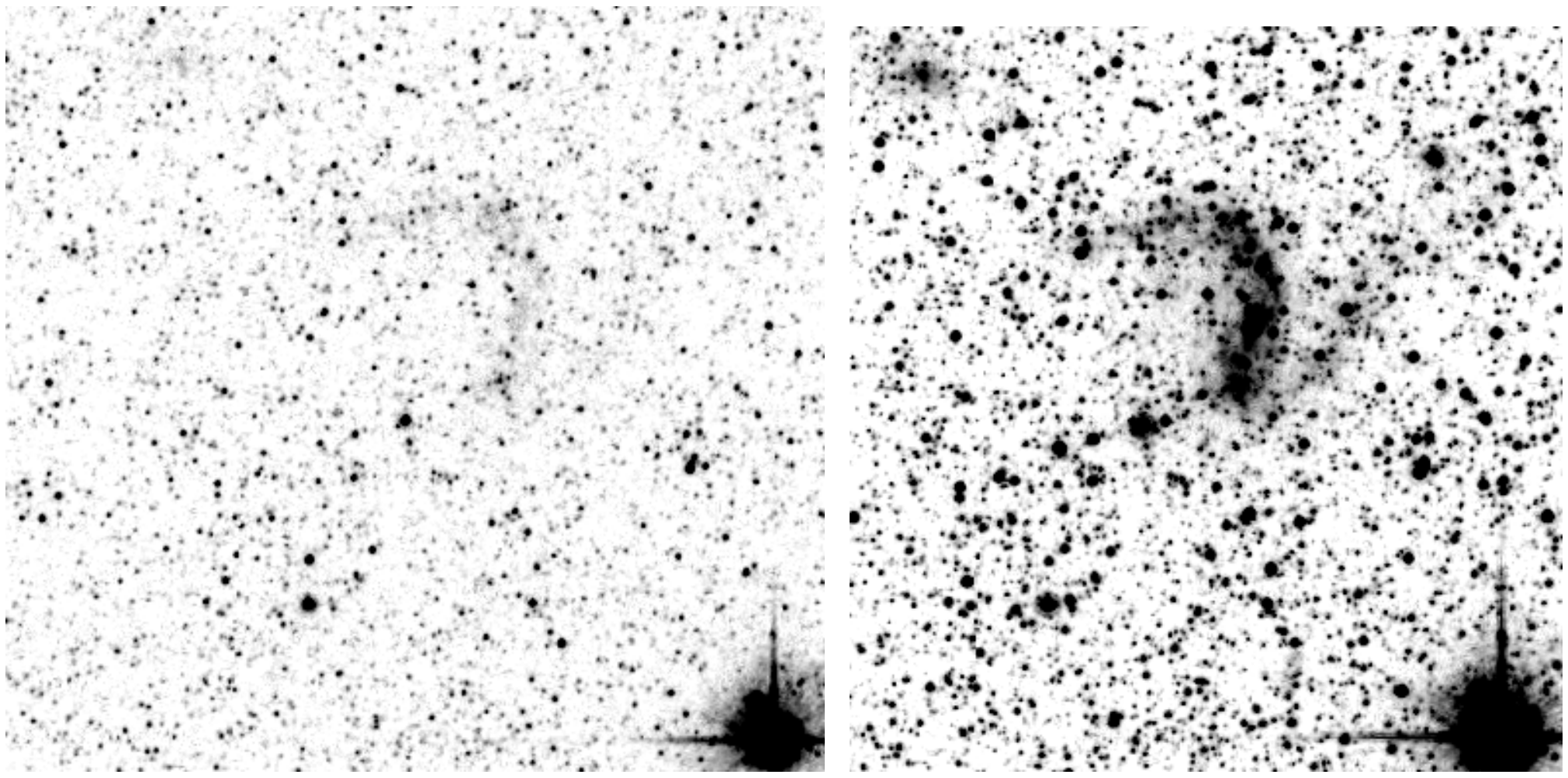

Figure 7. This pair of images shows Galaxy no. 356, a reddened SBc galaxy that is barely visible on the V image (left), but conspicuous on the I image (right).

## 5. FURTHER ANALYSIS

We are making this data set available for persons who might be interested in using the galaxies for analysis of the optical properties of the disk. For the brighter galaxies, this might be possible using multi-aperture spectroscopy. There are a number of ways to use these kinds of data (Section 1), but we have found that, except for gross statistical results, it is not possible to do much detailed analysis without measuring the radial velocities of the galaxies. We do not have the facilities or the fortitude to carry out such an observational program at this time.

We are grateful to the Space Telescope Science Institute for making the Hubble images publically available.

TABLE 1. GALAXY CANDIDATES WITH V, I PHOTOMETRY

| No. | V | V err | I | I err | RA | Dec | a arcsec |
|---|---|---|---|---|---|---|---|
| 1 | 22.79 | 0.08 | 20.55 | 0.05 | 9.49127 | 39.64722 | 0.15 |
| 2 | 25.15 | 0.22 | 22.85 | 0.13 | 9.49198 | 39.64924 | |
| 3 | 23.67 | 0.26 | 21.93 | 0.33 | 9.49201 | 39.64727 | 0.24 |
| 4 | 24.78 | 0.16 | 22.63 | 0.11 | 9.49225 | 39.63477 | 0.23 |
| 5 | 24.1 | 0.17 | 22.44 | 0.16 | 9.49254 | 39.61357 | |
| 6 | 24.38 | 0.17 | 23.76 | 0.21 | 9.49257 | 39.62200 | |
| 7 | 26.51 | 0.37 | 24.54 | 0.28 | 9.49273 | 39.65116 | 0.15 |
| 8 | 22.97 | 0.08 | 20.77 | 0.06 | 9.49301 | 39.63475 | 0.16 |
| 9 | 26.39 | 0.34 | 24.65 | 0.29 | 9.49308 | 39.64287 | 0.11 |
| 10 | 26.52 | 0.52 | 26.44 | 0.62 | 9.49332 | 39.62490 | |
| 11 | 22.69 | 0.08 | 20.65 | 0.06 | 9.49336 | 39.63174 | 0.26 |
| 12 | 24.34 | 0.18 | 22.94 | 0.16 | 9.49340 | 39.61217 | |
| 13 | 26.06 | 0.31 | 24.97 | 0.32 | 9.49359 | 39.65572 | |
| 14 | 25.42 | 0.37 | 24.15 | 0.5 | 9.49374 | 39.60584 | 0.32 |
| 15 | 25.63 | 0.28 | 25.12 | 0.34 | 9.49379 | 39.65267 | 0.20 |
| 16 | 26.44 | 0.53 | 28.03 | 0.85 | 9.49380 | 39.61921 | 0.18 |
| 17 | 26.97 | 0.6 | 25.52 | 0.45 | 9.49381 | 39.62432 | 0.34 |
| 18 | 24.05 | 0.16 | 22.83 | 0.16 | 9.49450 | 39.64689 | 0.79 |
| 19 | 25.92 | 0.72 | 23.88 | 0.59 | 9.49456 | 39.64189 | 0.24 |
| 20 | 24.39 | 0.18 | 23.87 | 0.24 | 9.49510 | 39.61738 | 0.28 |
| 21 | 24.79 | 0.18 | 22.91 | 0.13 | 9.49525 | 39.60694 | 0.22 |
| 22 | 23.75 | 0.14 | 22.69 | 0.14 | 9.49542 | 39.62426 | 0.69 |
| 23 | 25.98 | 0.69 | 26.12 | 0.75 | 9.49570 | 39.62978 | 0.25 |
| 24 | 23.72 | 0.1 | 21.55 | 0.07 | 9.49573 | 39.63664 | 0.11 |
| 25 | 25.16 | 0.22 | 23.77 | 0.2 | 9.49575 | 39.63187 | |
| 26 | 21.2 | 0.06 | 20.04 | 0.07 | 9.49582 | 39.64659 | 0.20 |
| 27 | 24.91 | 0.21 | 23.22 | 0.16 | 9.49587 | 39.63926 | 0.31 |
| 28 | 25.67 | 0.27 | 23.24 | 0.15 | 9.49598 | 39.62532 | 0.23 |
| 29 | 23.89 | 0.18 | 22.32 | 0.15 | 9.49630 | 39.64972 | 0.00 |
| 30 | 24.32 | 0.17 | 23.79 | 0.22 | 9.49723 | 39.61545 | 0.00 |
| 31 | 24.73 | 0.23 | 23.49 | 0.38 | 9.49726 | 39.65341 | 0.19 |
| 32 | 29.69 | 1.48 | 25.16 | 0.31 | 9.49760 | 39.60546 | 0.65 |
| 33 | 26.15 | 0.34 | 25.55 | 0.41 | 9.49762 | 39.65326 | 0.23 |
| 34 | 23.59 | 0.1 | 21.29 | 0.07 | 9.49780 | 39.63905 | 0.31 |
| 35 | 24.98 | 0.19 | 23.46 | 0.17 | 9.49784 | 39.61272 | 0.31 |
| 36 | 26.86 | 0.42 | 24.64 | 0.3 | 9.49793 | 39.63754 | 0.78 |
| 37 | 23.69 | 0.13 | 22.59 | 0.12 | 9.49794 | 39.61634 | 0.59 |
| 38 | 24.55 | 0.14 | 23.18 | 0.14 | 9.49851 | 39.60584 | 0.36 |
| 39 | 26.57 | 0.4 | 25.26 | 0.36 | 9.49910 | 39.65506 | 0.27 |
| 40 | 25.62 | 0.33 | 25.54 | 0.37 | 9.49935 | 39.65662 | 0.26 |
| 41 | 27.29 | 0.84 | 25.82 | 0.57 | 9.49964 | 39.60562 | 0.41 |
| 42 | 25.07 | 0.27 | 23.01 | 0.17 | 9.49984 | 39.64438 | 0.62 |
| 43 | 25.31 | 0.31 | 24.33 | 0.29 | 9.50000 | 39.63400 | 0.31 |
| 44 | 25.48 | 0.32 | 22.87 | 0.18 | 9.50039 | 39.61159 | 0.24 |
| 45 | 25.32 | 0.28 | 24.78 | 0.33 | 9.50050 | 39.60906 | 0.26 |
| 46 | 24.3 | 0.16 | 23.53 | 0.19 | 9.50071 | 39.61298 | 1.56 |
| 47 | 25.31 | 0.24 | 24.8 | 0.3 | 9.50073 | 39.65555 | 0.00 |
| 48 | 26.16 | 0.34 | 24.75 | 0.29 | 9.50106 | 39.63440 | 0.00 |
| 49 | 24.11 | 0.14 | 23.09 | 0.14 | 9.50113 | 39.62278 | 0.15 |
| 50 | 25.27 | 0.35 | 24.21 | 0.38 | 9.50114 | 39.64664 | 0.26 |
| 51 | 25.4 | 0.24 | 24.42 | 0.27 | 9.50119 | 39.63718 | 0.33 |
| 52 | 25.35 | 0.24 | 24.49 | 0.26 | 9.50166 | 39.61634 | 0.18 |
| 53 | 26.45 | 0.38 | 25.71 | 0.43 | 9.50167 | 39.64589 | 0.33 |
| 54 | 25.75 | 0.37 | 24.55 | 0.29 | 9.50172 | 39.61759 | 0.37 |
| 55 | 24.81 | 0.17 | 22.99 | 0.13 | 9.50206 | 39.65227 | 0.09 |
| 56 | 19.91 | 0.03 | 18.15 | 0.02 | 9.50216 | 39.62423 | 0.71 |
| 57 | 24.95 | 0.21 | 24.47 | 0.26 | 9.50229 | 39.64596 | 0.00 |
| 58 | 23.27 | 0.13 | 22.35 | 0.13 | 9.50269 | 39.60491 | 0.32 |
| 59 | 24.81 | 0.2 | 23.5 | 0.19 | 9.50281 | 39.63165 | 0.93 |
| 60 | 21.92 | 0.11 | 20.39 | 0.09 | 9.50336 | 39.63258 | 0.24 |
| 61 | 27.18 | 0.66 | 25.32 | 0.49 | 9.50411 | 39.63285 | 0.21 |
| 62 | 25.15 | 0.29 | 23.92 | 0.26 | 9.50415 | 39.64698 | 0.23 |
| 63 | 24.13 | 0.12 | 22.46 | 0.1 | 9.50421 | 39.62687 | 0.49 |
| 64 | 25.62 | 0.48 | 24.71 | 0.5 | 9.50436 | 39.62534 | 0.56 |
| 65 | 23.36 | 0.12 | 22.03 | 0.13 | 9.50501 | 39.62920 | 0.18 |
| 66 | 27.22 | 0.62 | 25.28 | 0.41 | 9.50506 | 39.63431 | 0.29 |
| 67 | 20.34 | 0.04 | 19.3 | 0.05 | 9.50534 | 39.62962 | 0.38 |
| 68 | 21.47 | 0.05 | 20.33 | 0.05 | 9.50587 | 39.61326 | 1.12 |
| 69 | 26.89 | 0.47 | 25.71 | 0.46 | 9.50602 | 39.62453 | 1.20 |
| 70 | 25.64 | 0.31 | 24.69 | 0.5 | 9.50633 | 39.62129 | 0.54 |
| 71 | 22.87 | 0.13 | 20.61 | 0.05 | 9.85409 | 40.77928 | 0.22 |
| 72 | 23.63 | 0.22 | 20.96 | 0.11 | 9.86495 | 40.78593 | 0.69 |
| 73 | 24.33 | 0.24 | 21.24 | 0.06 | 9.86497 | 40.78594 | 0.25 |
| 74 | 23.71 | 0.23 | 21.29 | 0.14 | 9.86553 | 40.77414 | 0.11 |
| 75 | 24.3 | 0.27 | 21.43 | 0.08 | 9.86555 | 40.77411 | 0.00 |
| 76 | 24.01 | 0.24 | 21.39 | 0.08 | 9.87533 | 40.76306 | 0.20 |
| 77 | 24.01 | 0.25 | 21.71 | 0.09 | 9.87695 | 40.78250 | 0.00 |
| 78 | 24.12 | 0.3 | 21.75 | 0.13 | 9.87696 | 40.78248 | 0.12 |
| 79 | 23.59 | 0.18 | 21.64 | 0.13 | 9.87796 | 40.76684 | 0.31 |
| 80 | 23.66 | 0.22 | 21.7 | 0.15 | 9.87796 | 40.76681 | 0.23 |
| 81 | 23.67 | 0.48 | 21.12 | 0.14 | 9.87891 | 40.77470 | 0.22 |
| 82 | 23.67 | 0.23 | 21.15 | 0.13 | 9.87892 | 40.77466 | 0.00 |
| 83 | 20.94 | 0.06 | 19.36 | 0.04 | 9.88306 | 40.76368 | 0.00 |
| 84 | 20.88 | 0.06 | 19.32 | 0.05 | 9.88307 | 40.76370 | 0.19 |
| 85 | 21.41 | 0.13 | 19.39 | 0.1 | 9.89448 | 40.76067 | 0.65 |
| 86 | 17.97 | 0.03 | 16.49 | 0.02 | 9.89668 | 40.80449 | 0.19 |
| 87 | 24.66 | 0.48 | 22.52 | 0.61 | 9.90002 | 40.76598 | 0.31 |
| 88 | 21.39 | 0.07 | 20.04 | 0.05 | 9.90439 | 40.77083 | 0.36 |
| 89 | 22.48 | 0.11 | 21.16 | 0.09 | 9.91077 | 40.78892 | 0.27 |
| 90 | 21.45 | 0.07 | 20.12 | 0.08 | 9.91080 | 40.78893 | 0.26 |
| 91 | 22.78 | 0.12 | 21.21 | 0.06 | 9.92681 | 40.77806 | 0.41 |
| 92 | 22.78 | 0.16 | 21.57 | 0.19 | 10.29634 | 40.64711 | 0.93 |
| 93 | 22.57 | 0.13 | 21.47 | 0.19 | 10.30242 | 40.62460 | 0.24 |
| 94 | 20.79 | 0.06 | 20.1 | 0.06 | 10.30450 | 40.64760 | 0.21 |
| 95 | 23.96 | 0.2 | 20.81 | 0.07 | 10.30491 | 40.63956 | 0.18 |
| 96 | 21.11 | 0.06 | 19.73 | 0.07 | 10.30654 | 40.65182 | 0.23 |
| 97 | 21.66 | 0.08 | 20.33 | 0.07 | 10.30676 | 40.65518 | 0.49 |
| 98 | 23.78 | 0.17 | 24.89 | 0.28 | 10.31047 | 40.63501 | 0.56 |
| 99 | 22.93 | 0.14 | 21.35 | 0.12 | 10.31185 | 40.66238 | 0.18 |
| 100 | 23.13 | 0.23 | 20.78 | 0.09 | 10.31799 | 40.62007 | 0.29 |
| 101 | 22.78 | 0.13 | 21.22 | 0.06 | 10.33409 | 40.66811 | 0.38 |
| 102 | 21.37 | 0.08 | 19.96 | 0.08 | 10.34175 | 40.65849 | 1.12 |
| 103 | 20.65 | 0.08 | 19.93 | 0.15 | 10.34238 | 40.63564 | 1.20 |
| 104 | 23.83 | 0.23 | 21.3 | 0.09 | 10.34478 | 40.62476 | 0.54 |
| 105 | 20.74 | 0.07 | 19.35 | 0.06 | 10.34621 | 40.65591 | 0.31 |
| 106 | 20.98 | 0.07 | 20.35 | 0.12 | 10.35254 | 40.62846 | 0.22 |
| 107 | 24.8 | 0.48 | 21.63 | 0.22 | 10.35693 | 40.65880 | 0.32 |
| 108 | 20.34 | 0.06 | 19.03 | 0.04 | 10.46145 | 41.41131 | 0.10 |
| 109 | 18.56 | 0.09 | 17.12 | 0.03 | 10.46384 | 41.41642 | 0.00 |
| 110 | 17.61 | 0.04 | 16.3 | 0.03 | 10.46558 | 41.41164 | 0.25 |
| 111 | 23.74 | 0.23 | 22.28 | 0.22 | 10.46879 | 41.40725 | 0.29 |
| 112 | 20.83 | 0.07 | 19.53 | 0.08 | 10.47101 | 41.42810 | 0.27 |
| 113 | 22.55 | 0.13 | 21.93 | 0.24 | 10.47327 | 41.42727 | 0.40 |
| 114 | 21.79 | 0.1 | 21.14 | 0.15 | 10.47471 | 41.42596 | 0.34 |
| 115 | 21.37 | 0.08 | 19.76 | 0.06 | 10.47886 | 41.44723 | 0.31 |
| 116 | 19.67 | 0.07 | 17.96 | 0.04 | 10.48469 | 41.42367 | 0.29 |
| 117 | 22.54 | 0.13 | 21.34 | 0.11 | 10.48612 | 41.42047 | 0.33 |
| 118 | 22.7 | 0.14 | 20.57 | 0.08 | 10.48733 | 41.43409 | 0.32 |
| 119 | 21.92 | 0.12 | 20.2 | 0.11 | 10.48961 | 41.43236 | 0.98 |
| 120 | 21.51 | 0.09 | 20.64 | 0.11 | 10.49118 | 41.43910 | 0.56 |
| 121 | 21.54 | 0.22 | 20.48 | 0.05 | 10.49189 | 41.39958 | 0.34 |
| 122 | 22.2 | 0.12 | 20.44 | 0.07 | 10.50530 | 41.42532 | 0.53 |
| 123 | 18.35 | 0.03 | 17.03 | 0.03 | 10.50663 | 41.41753 | 0.30 |
| 124 | 21.47 | 0.07 | 20.36 | 0.09 | 10.50888 | 41.43442 | 0.30 |
| 125 | 20.5 | 0.23 | 18.17 | 0.07 | 10.51057 | 41.41026 | 0.00 |
| 126 | 23.17 | 0.14 | 20.96 | 0.05 | 10.52450 | 41.40930 | 0.00 |
| 127 | 18.94 | 0.09 | 17.55 | 0.11 | 10.52896 | 41.05359 | 0.23 |
| 128 | 21.78 | 0.07 | 20.49 | 0.07 | 10.53141 | 41.01995 | 0.49 |
| 129 | 19.24 | 0.02 | 17.79 | 0.02 | 10.53237 | 41.01936 | 0.56 |
| 130 | 20.59 | 0.08 | 19.15 | 0.09 | 10.53305 | 41.02007 | 0.18 |
| 131 | 24.93 | 0.38 | 22.23 | 0.09 | 10.53895 | 41.42230 | 0.00 |
| 132 | 21.04 | 0.14 | 20.3 | 0.14 | 10.54105 | 41.42819 | 0.00 |
| 133 | 22.82 | 0.21 | 20.74 | 0.07 | 10.54182 | 41.50808 | 0.00 |
| 134 | 20.44 | 0.05 | 18.98 | 0.06 | 10.54338 | 41.01758 | 1.20 |
| 135 | 23.31 | 0.17 | 20.96 | 0.1 | 10.54340 | 41.51046 | 0.00 |
| 136 | 19.9 | 0.03 | 18.37 | 0.04 | 10.54465 | 41.05216 | 0.31 |
| 137 | 21.5 | 0.14 | 20.78 | 0.18 | 10.54619 | 41.50936 | 0.00 |
| 138 | 22.57 | 0.21 | 21.08 | 0.21 | 10.54962 | 41.56207 | 0.39 |
| 139 | 19.97 | 0.03 | 18.46 | 0.04 | 10.55128 | 41.05401 | 0.00 |
| 140 | 21.52 | 0.12 | 19.68 | 0.08 | 10.55154 | 41.50059 | 0.15 |
| 141 | 23.61 | 0.19 | 22.38 | 0.14 | 10.55673 | 40.94584 | 0.53 |
| 142 | 21.39 | 0.11 | 19.67 | 0.07 | 10.55679 | 40.92759 | 0.30 |
| 143 | 21.73 | 0.07 | 19.73 | 0.08 | 10.55751 | 41.01658 | 0.23 |
| 144 | 22.22 | 0.13 | 20.97 | 0.11 | 10.55833 | 41.51323 | 0.27 |
| 145 | 24.01 | 0.17 | 22.3 | 0.13 | 10.56025 | 41.55902 | 0.00 |
| 146 | 24.53 | 0.24 | 22.92 | 0.18 | 10.56337 | 41.56166 | 0.31 |
| 147 | 23.4 | 0.27 | 21.64 | 0.12 | 10.56370 | 41.49948 | 0.35 |
| 148 | 21.49 | 0.06 | 19.59 | 0.07 | 10.56391 | 41.55874 | 0.27 |
| 149 | 20.83 | 0.09 | 19.51 | 0.09 | 10.56446 | 40.95871 | 0.00 |
| 150 | 21.71 | 0.09 | 20.5 | 0.08 | 10.56470 | 41.50959 | 0.28 |
| 151 | 23.84 | 0.16 | 21.35 | 0.1 | 10.56511 | 41.54190 | 0.56 |
| 152 | 24.73 | 0.4 | 22.82 | 0.21 | 10.56689 | 41.39301 | 0.00 |
| 153 | 21.63 | 0.08 | 19.91 | 0.09 | 10.56944 | 41.02521 | 0.23 |
| 154 | 24.76 | 0.22 | 23.25 | 0.32 | 10.57060 | 41.54258 | |
| 155 | 22.34 | 0.07 | 20.95 | 0.06 | 10.57068 | 41.57755 | 0.98 |
| 156 | 24.29 | 0.14 | 21.9 | 0.1 | 10.57221 | 41.56829 | 0.56 |
| 157 | 23.11 | 0.09 | 21.61 | 0.08 | 10.57261 | 41.56777 | 0.34 |
| 158 | 23.36 | 0.2 | 22.47 | 0.28 | 10.57262 | 40.92711 | 0.20 |
| 159 | 23.97 | 0.26 | 22.59 | 0.29 | 10.57309 | 41.02828 | 0.29 |
| 160 | 19.67 | 0.07 | 18.67 | 0.07 | 10.57326 | 40.92102 | 0.20 |
| 161 | 21.65 | 0.07 | 20.25 | 0.04 | 10.57340 | 40.98114 | 0.34 |
| 162 | 22.17 | 0.11 | 20.86 | 0.11 | 10.57375 | 40.94186 | 0.35 |
| 163 | 21.6 | 0.07 | 19.78 | 0.07 | 10.57412 | 41.55529 | 0.00 |
| 164 | 23.59 | 0.42 | 22.74 | 0.23 | 10.57436 | 41.53836 | 0.15 |
| 165 | 20.77 | 0.06 | 19.7 | 0.05 | 10.57620 | 41.48832 | 0.15 |
| 166 | 21.22 | 0.04 | 19.2 | 0.07 | 10.57648 | 41.56697 | 0.50 |
| 167 | 21.86 | 0.13 | 19.92 | 0.09 | 10.57724 | 40.92801 | 0.42 |
| 168 | 23.74 | 0.18 | 21.6 | 0.13 | 10.57823 | 40.95994 | 0.22 |
| 169 | 22.75 | 0.17 | 20.76 | 0.11 | 10.57870 | 40.92144 | 0.41 |
| 170 | 22.14 | 0.08 | 20.38 | 0.08 | 10.57909 | 41.56391 | 0.23 |
| 171 | 22.62 | 0.32 | 20.61 | 0.14 | 10.57985 | 40.95678 | 0.35 |
| 172 | 22.21 | 0.08 | 21.2 | 0.09 | 10.58212 | 41.57483 | 0.15 |
| 173 | 22.1 | 0.18 | 20.42 | 0.1 | 10.58373 | 41.48489 | 0.82 |
| 174 | 20.16 | 0.04 | 18.36 | 0.04 | 10.58546 | 41.04691 | 0.00 |
| 175 | 22.56 | 0.18 | 20.64 | 0.06 | 10.58584 | 40.92114 | 0.00 |
| 176 | 20.89 | 0.07 | 18.93 | 0.03 | 10.58639 | 40.95382 | 0.22 |
| 177 | 21.79 | 0.07 | 20.46 | 0.09 | 10.58744 | 41.01458 | 0.00 |
| 178 | 20.51 | 0.07 | 19.12 | 0.07 | 10.58776 | 41.01521 | 0.00 |
| 179 | 21.79 | 0.07 | 21.07 | 0.09 | 10.58833 | 41.01630 | 0.00 |
| 180 | 20.79 | 0.1 | 19.59 | 0.06 | 10.58955 | 41.50012 | 0.45 |
| 181 | 21.91 | 0.12 | 20.41 | 0.09 | 10.58979 | 40.96729 | 0.14 |
| 182 | 18.78 | 0.05 | 18.19 | 0.05 | 10.59060 | 41.36818 | 0.00 |
| 183 | 21.52 | 0.1 | 20.84 | 0.09 | 10.59061 | 41.41070 | 0.60 |
| 184 | 23.13 | 0.11 | 22.13 | 0.12 | 10.59075 | 41.57539 | 0.41 |
| 185 | 23.55 | 0.18 | 22.12 | 0.13 | 10.59250 | 40.93503 | 0.29 |
| 186 | 24.18 | 0.25 | 21.71 | 0.12 | 10.59280 | 41.05426 | 0.22 |
| 187 | 20.61 | 0.04 | 19.74 | 0.04 | 10.59373 | 41.41922 | 0.39 |
| 188 | 20.03 | 0.06 | 18.28 | 0.05 | 10.59516 | 40.95309 | 0.82 |
| 189 | 21.33 | 0.07 | 20.27 | 0.09 | 10.59638 | 41.03022 | 0.20 |
| 190 | 17.06 | 0.02 | 15.82 | 0.02 | 10.59663 | 41.36214 | 0.39 |
| 191 | 24.09 | 0.19 | 22.44 | 0.18 | 10.59671 | 41.57425 | 0.12 |
| 192 | 22.61 | 0.15 | 21.31 | 0.08 | 10.59699 | 41.40218 | 0.53 |
| 193 | 22 | 0.13 | 21.01 | 0.13 | 10.59813 | 41.41070 | 0.23 |
| 194 | 22.12 | 0.12 | 21.28 | 0.17 | 10.59918 | 40.95969 | 0.84 |
| 195 | 23.83 | 0.26 | 21.39 | 0.16 | 10.60227 | 41.41582 | 0.31 |
| 196 | 22.27 | 0.13 | 20.56 | 0.1 | 10.60265 | 41.52361 | 0.38 |
| 197 | 23.32 | 0.14 | 21.56 | 0.14 | 10.60370 | 41.53855 | 0.15 |
| 198 | 23.69 | 0.26 | 21.1 | 0.12 | 10.60509 | 41.50018 | 0.53 |
| 199 | 21.61 | 0.05 | 20.47 | 0.07 | 10.60679 | 41.56826 | 0.29 |
| 200 | 19.83 | 0.05 | 18.86 | 0.05 | 10.61122 | 41.38658 | 0.59 |
| 201 | 23.17 | 0.24 | 21.05 | 0.22 | 10.61366 | 41.57905 | 0.35 |
| 202 | 21.96 | 0.11 | 19.89 | 0.05 | 10.61522 | 41.41034 | 2.86 |
| 203 | 22.35 | 0.16 | 20.42 | 0.14 | 10.62015 | 41.40231 | 0.29 |
| 204 | 23.18 | 0.17 | 21.37 | 0.08 | 10.63529 | 41.59673 | 0.00 |
| 205 | 21.76 | 0.09 | 19.87 | 0.07 | 10.63800 | 41.59089 | 0.22 |
| 206 | 20.45 | 0.05 | 20.26 | 0.09 | 10.64370 | 41.04803 | 0.81 |
| 207 | 22.32 | 0.17 | 19.92 | 0.08 | 10.64730 | 41.62962 | 0.13 |
| 208 | 24.02 | 0.22 | 22.72 | 0.14 | 10.64743 | 41.59663 | 1.76 |
| 209 | 23.16 | 0.54 | 22.21 | 0.57 | 10.64798 | 41.62654 | 0.38 |
| 210 | 24.15 | 0.24 | 23.47 | 0.25 | 10.65770 | 41.59286 | 0.00 |
| 211 | 24.85 | 0.34 | 22.38 | 0.18 | 10.65791 | 41.04842 | 0.21 |
| 212 | 22.47 | 0.12 | 21.44 | 0.12 | 10.65930 | 41.58808 | 0.60 |
| 213 | 24.39 | 0.48 | 21.43 | 0.1 | 10.66057 | 41.06393 | 0.00 |
| 214 | 22.93 | 0.17 | 20.72 | 0.06 | 10.66202 | 41.58023 | 0.58 |
| 215 | 25.53 | 0.46 | 23.32 | 0.33 | 10.66486 | 41.63193 | 0.36 |
| 216 | 23.66 | 0.19 | 22.28 | 0.13 | 10.66498 | 41.63028 | 0.56 |
| 217 | 18.57 | 0.04 | 17.05 | 0.04 | 10.66998 | 41.47275 | 0.00 |
| 218 | 22.42 | 0.1 | 22.53 | 0.16 | 10.67079 | 41.61169 | 0.53 |
| 219 | 23.95 | 0.19 | 22.41 | 0.16 | 10.67637 | 41.58223 | 0.38 |
| 220 | 22.98 | 0.15 | 20.53 | 0.06 | 10.67933 | 41.60215 | 0.53 |
| 221 | 21.19 | 0.18 | 19.73 | 0.13 | 10.68092 | 41.59930 | 0.23 |
| 222 | 22.62 | 0.22 | 20.72 | 0.19 | 10.68248 | 41.60095 | 0.24 |
| 223 | 24.17 | 0.25 | 22.84 | 0.2 | 10.68333 | 41.47855 | 0.15 |
| 224 | 20.25 | 0.11 | 18.74 | 0.09 | 10.68416 | 41.47609 | 0.45 |
| 225 | 21.7 | 0.22 | 21.26 | 0.26 | 10.68544 | 41.52175 | 0.41 |
| 226 | 21.19 | 0.06 | 18.73 | 0.02 | 10.68868 | 41.59322 | 0.00 |
| 227 | 23.44 | 0.37 | 21.39 | 0.14 | 10.69030 | 41.51939 | 0.88 |
| 228 | 26.07 | 0.4 | 21.44 | 0.16 | 10.69647 | 41.44302 | 0.00 |
| 229 | 22.57 | 0.12 | 21.28 | 0.15 | 10.69721 | 41.47751 | 0.19 |
| 230 | 21.52 | 0.09 | 20.58 | 0.1 | 10.70082 | 41.60678 | 0.47 |
| 231 | 21.58 | 0.09 | 19.53 | 0.04 | 10.70381 | 41.69083 | 0.00 |
| 232 | 22.36 | 0.11 | 21.26 | 0.1 | 10.70463 | 41.52258 | 0.31 |
| 233 | 20.32 | 0.06 | 19.29 | 0.06 | 10.70540 | 41.45872 | 0.50 |
| 234 | 22.83 | 0.13 | 21.91 | 0.13 | 10.70639 | 41.53426 | 0.28 |
| 235 | 22.45 | 0.13 | 20.85 | 0.16 | 10.70860 | 41.12614 | 0.22 |
| 236 | 20.99 | 0.07 | 20.03 | 0.07 | 10.71936 | 41.48632 | 0.22 |
| 237 | 22.25 | 0.11 | 21.78 | 0.12 | 10.72182 | 41.54807 | 0.21 |
| 238 | 20.08 | 0.07 | 18.68 | 0.07 | 10.72434 | 41.53513 | 0.27 |
| 239 | 22.7 | 0.15 | 21.18 | 0.13 | 10.72540 | 41.52810 | 0.00 |
| 240 | 21.76 | 0.07 | 21.04 | 0.06 | 10.72700 | 40.97586 | 0.30 |
| 241 | 19.84 | 0.06 | 18.54 | 0.06 | 10.72812 | 41.69889 | 0.12 |
| 242 | 25.25 | 0.38 | 23.32 | 0.21 | 10.72956 | 41.54202 | 0.31 |
| 243 | 22.65 | 0.16 | 20.85 | 0.07 | 10.72958 | 41.52674 | 0.20 |
| 244 | 25.24 | 0.47 | 21.5 | 0.1 | 10.73034 | 41.53589 | 0.15 |
| 245 | 24.36 | 0.37 | 22.11 | 0.12 | 10.73178 | 41.71300 | 0.68 |
| 246 | 21.76 | 0.09 | 19.63 | 0.05 | 10.73232 | 41.68052 | 0.29 |
| 247 | 25.42 | 0.52 | 22.05 | 0.15 | 10.73334 | 40.97062 | 0.31 |
| 248 | 20.55 | 0.17 | 19.54 | 0.06 | 10.73406 | 41.15077 | 0.12 |
| 249 | 23.82 | 0.2 | 22.35 | 0.18 | 10.73723 | 41.70936 | 0.32 |
| 250 | 22.64 | 0.37 | 20.85 | 0.24 | 10.74080 | 40.96884 | 1.08 |
| 251 | 21.41 | 0.15 | 19.93 | 0.07 | 10.74237 | 41.70735 | 0.39 |
| 252 | 23.71 | 0.26 | 22.29 | 0.15 | 10.74293 | 41.69267 | 0.28 |
| 253 | 21.58 | 0.09 | 20.53 | 0.07 | 10.74447 | 41.70123 | 0.42 |
| 254 | 23.72 | 0.18 | 22.42 | 0.12 | 10.74594 | 41.49840 | 0.30 |
| 255 | 24.62 | 0.27 | 22.3 | 0.13 | 10.74607 | 41.68765 | 0.37 |
| 256 | 20.12 | 0.04 | 19.07 | 0.05 | 10.74730 | 41.50552 | 1.45 |
| 257 | 22.66 | 0.19 | 20.96 | 0.08 | 10.74877 | 41.01719 | 0.21 |
| 258 | 21.47 | 0.14 | 20.12 | 0.03 | 10.74968 | 41.50359 | 0.22 |
| 259 | 21.34 | 0.09 | 20.24 | 0.06 | 10.74997 | 41.00165 | 0.00 |
| 260 | 22.42 | 0.38 | 20.14 | 0.2 | 10.75144 | 40.96294 | 0.00 |
| 261 | 23.84 | 0.28 | 21.9 | 0.1 | 10.75261 | 41.66527 | 0.26 |
| 262 | 21.41 | 0.07 | 20.6 | 0.07 | 10.75298 | 41.02168 | 0.00 |
| 263 | 23.27 | 0.22 | 21.46 | 0.11 | 10.75343 | 41.68133 | 0.00 |
| 264 | 22.72 | 0.48 | 21.96 | 0.57 | 10.75373 | 40.95792 | 0.15 |
| 265 | 21.83 | 0.1 | 21.14 | 0.12 | 10.75461 | 41.52332 | 0.20 |
| 266 | 24.48 | 0.44 | 21.5 | 0.11 | 10.75469 | 41.67192 | 0.38 |
| 267 | 22.64 | 0.13 | 21.41 | 0.11 | 10.75565 | 41.70253 | 0.23 |
| 268 | 23.86 | 0.22 | 21.99 | 0.21 | 10.75786 | 41.14850 | 0.34 |
| 269 | 22.9 | 0.6 | 20.39 | 0.09 | 10.75870 | 40.95387 | 0.29 |
| 270 | 20.9 | 0.08 | 19.48 | 0.05 | 10.75959 | 41.00542 | 0.27 |
| 271 | 25.22 | 0.34 | 22.73 | 0.15 | 10.76030 | 41.67368 | 0.19 |
| 272 | 21.92 | 0.13 | 20.18 | 0.1 | 10.76070 | 40.98603 | 0.32 |
| 273 | 21.92 | 0.29 | 19.96 | 0.13 | 10.76168 | 40.95623 | 0.28 |
| 274 | 20.52 | 0.09 | 19.53 | 0.08 | 10.76201 | 41.71405 | 0.12 |
| 275 | 25.24 | 0.37 | 22.1 | 0.1 | 10.76210 | 41.01634 | 0.28 |
| 276 | 24.58 | 0.63 | 21.91 | 0.11 | 10.76251 | 41.01560 | 0.42 |
| 277 | 24.02 | 0.21 | 22.37 | 0.13 | 10.76297 | 40.95702 | 0.20 |
| 278 | 24.4 | 0.25 | 23.22 | 0.16 | 10.76351 | 41.70684 | 0.00 |
| 279 | 24.18 | 0.28 | 22.68 | 0.17 | 10.76436 | 40.96194 | 0.26 |
| 280 | 22.7 | 0.18 | 21.9 | 0.14 | 10.76507 | 40.96516 | 0.18 |
| 281 | 24.8 | 0.28 | 23.03 | 0.16 | 10.76531 | 40.96921 | 0.34 |
| 282 | 19.63 | 0.08 | 17.96 | 0.06 | 10.76685 | 40.96825 | 1.54 |
| 283 | 21.67 | 0.08 | 20.68 | 0.07 | 10.76712 | 41.51911 | 0.28 |
| 284 | 24.57 | 0.27 | 23.2 | 0.22 | 10.76791 | 41.11417 | 0.19 |
| 285 | 21.91 | 0.15 | 20.32 | 0.11 | 10.77098 | 41.01171 | 0.00 |
| 286 | 23.53 | 0.27 | 21.36 | 0.11 | 10.77215 | 41.12236 | 0.15 |
| 287 | 26.02 | 0.55 | 24.16 | 0.29 | 10.77310 | 41.71563 | 0.20 |
| 288 | 23.41 | 0.2 | 21.38 | 0.12 | 10.77510 | 41.13543 | 0.15 |
| 289 | 22.27 | 0.11 | 20.63 | 0.1 | 10.77765 | 41.52708 | 0.26 |
| 290 | 21.63 | 0.17 | 19.53 | 0.05 | 10.77769 | 40.99062 | 0.23 |
| 291 | 23.97 | 0.22 | 22.43 | 0.16 | 10.77987 | 40.97913 | 0.22 |
| 292 | 23.13 | 0.19 | 21.82 | 0.14 | 10.78447 | 41.00143 | 0.93 |
| 293 | 21.95 | 0.16 | 20.25 | 0.13 | 10.78738 | 40.99355 | 0.23 |
| 294 | 22.03 | 0.18 | 20.37 | 0.06 | 10.79036 | 40.98719 | 0.19 |
| 295 | 22.95 | 0.13 | 22.02 | 0.12 | 10.79112 | 40.99275 | 0.29 |
| 296 | 23.52 | 0.25 | 21.74 | 0.21 | 10.79532 | 40.98395 | 0.77 |
| 297 | 22.37 | 0.17 | 19.14 | 0.02 | 10.79591 | 40.98770 | 0.24 |
| 298 | 18.93 | 0.03 | 16.9 | 0.01 | 10.82524 | 41.67231 | 0.00 |
| 299 | 20.39 | 0.05 | 19.82 | 0.09 | 10.83467 | 41.64708 | 0.00 |
| 300 | 22.38 | 0.13 | 21.06 | 0.12 | 10.83499 | 41.67474 | 0.00 |
| 301 | 20.66 | 0.09 | 19.8 | 0.09 | 10.84000 | 41.67345 | 0.34 |
| 302 | 22.9 | 0.25 | 20.43 | 0.12 | 10.84420 | 41.63790 | 0.38 |
| 303 | 20.2 | 0.04 | 18.47 | 0.03 | 10.84598 | 41.04026 | 0.00 |
| 304 | 21.2 | 0.09 | 20.47 | 0.1 | 10.84684 | 41.67956 | 0.22 |
| 305 | 22.6 | 0.13 | 20.83 | 0.11 | 10.84714 | 41.08167 | 0.12 |
| 306 | 22.99 | 0.16 | 21.48 | 0.12 | 10.84891 | 41.07640 | 0.00 |
| 307 | 24.02 | 0.17 | 21.92 | 0.1 | 10.85304 | 41.06125 | 0.00 |
| 308 | 23.62 | 0.2 | 21.84 | 0.13 | 10.85799 | 41.64690 | 0.29 |
| 309 | 19.02 | 0.01 | 18.74 | 0.02 | 10.85934 | 41.08577 | 0.26 |
| 310 | 20.19 | 0.05 | 18.82 | 0.06 | 10.86153 | 41.06192 | 0.28 |
| 311 | 22.29 | 0.13 | 20.47 | 0.09 | 10.86965 | 41.65402 | 0.43 |
| 312 | 24.93 | 0.18 | 22.06 | 0.12 | 10.86986 | 41.04089 | 0.22 |
| 313 | 23.26 | 0.16 | 22.09 | 0.23 | 10.87045 | 41.07315 | 0.82 |
| 314 | 22.51 | 0.1 | 21.01 | 0.11 | 10.87399 | 41.07518 | 0.29 |
| 315 | 24.09 | 0.21 | 23.12 | 0.25 | 10.87439 | 41.05211 | 0.18 |
| 316 | 21.79 | 0.09 | 20.5 | 0.11 | 10.88062 | 41.05875 | 0.35 |
| 317 | 23.22 | 0.18 | 22.15 | 0.16 | 10.88165 | 41.04110 | 0.36 |
| 318 | 21.13 | 0.09 | 19.23 | 0.07 | 10.88521 | 41.07121 | 0.24 |
| 319 | 22.98 | 0.07 | 21.14 | 0.05 | 10.88750 | 41.03824 | 0.30 |
| 320 | 24.58 | 0.25 | 22.35 | 0.12 | 10.89097 | 41.05190 | 0.61 |
| 321 | 21.68 | 0.07 | 20.92 | 0.05 | 10.89103 | 41.65364 | 0.00 |
| 322 | 23.45 | 0.12 | 21.74 | 0.12 | 10.89501 | 41.04226 | 0.28 |
| 323 | 16.61 | 0.04 | 16.05 | 0.06 | 10.89541 | 41.16470 | 0.36 |
| 324 | 23.36 | 0.15 | 21.98 | 0.13 | 10.89568 | 41.67271 | 0.26 |
| 325 | 22.26 | 0.09 | 20.26 | 0.06 | 10.89802 | 41.04224 | 0.89 |
| 326 | 24.88 | 0.4 | 22.34 | 0.22 | 10.91003 | 41.73290 | 0.26 |
| 327 | 21.34 | 0.07 | 19.86 | 0.06 | 10.91004 | 41.71885 | 0.28 |
| 328 | 27.17 | 0.81 | 22.9 | 0.18 | 10.91140 | 41.73496 | 0.20 |
| 329 | 20.97 | 0.07 | 20.61 | 0.15 | 10.91433 | 41.16636 | 0.69 |
| 330 | 24.12 | 0.3 | 21.49 | 0.12 | 10.91714 | 41.15956 | 0.62 |
| 331 | 21.3 | 0.11 | 20.43 | 0.11 | 10.91936 | 41.19837 | 0.22 |
| 332 | 21.3 | 0.2 | 20.92 | 0.17 | 10.92185 | 41.75065 | 0.18 |
| 333 | 22.6 | 0.13 | 20.66 | 0.17 | 10.92246 | 41.75607 | 0.35 |
| 334 | 20.79 | 0.06 | 18.53 | 0.03 | 10.92485 | 41.73431 | 0.36 |
| 335 | 22.63 | 0.11 | 20.46 | 0.04 | 10.92563 | 41.20472 | 0.19 |
| 336 | 22.67 | 0.14 | 20.08 | 0.06 | 10.92858 | 41.14825 | 0.23 |
| 337 | 21.17 | 0.08 | 20.44 | 0.11 | 10.93049 | 41.15043 | 0.88 |
| 338 | 21.39 | 0.08 | 20.4 | 0.06 | 10.93083 | 41.73758 | 0.68 |
| 339 | 22.99 | 0.19 | 22.37 | 0.47 | 10.93135 | 41.13977 | 0.91 |
| 340 | 21.44 | 0.08 | 20.12 | 0.1 | 10.93299 | 41.19345 | 0.30 |
| 341 | 27.7 | 0.72 | 23.15 | 0.64 | 10.93858 | 41.13349 | 0.26 |
| 342 | 20.58 | 0.07 | 19.44 | 0.07 | 10.93947 | 41.14906 | 0.35 |
| 343 | 22.31 | 0.13 | 21.21 | 0.15 | 10.94166 | 41.19177 | 0.34 |
| 344 | 21.18 | 0.06 | 19.59 | 0.04 | 10.94246 | 41.17854 | 0.19 |
| 345 | 22.82 | 0.19 | 21.95 | 0.27 | 10.94565 | 41.17265 | 0.33 |
| 346 | 20.8 | 0.06 | 20.67 | 0.06 | 10.94888 | 41.16538 | 0.32 |
| 347 | 21.12 | 0.09 | 19.66 | 0.05 | 10.94979 | 41.15118 | 0.36 |
| 348 | 20.93 | 0.07 | 19.71 | 0.07 | 10.95036 | 41.19248 | 0.47 |
| 349 | 24.22 | 0.33 | 21.19 | 0.08 | 10.95137 | 41.74909 | 0.41 |
| 350 | 22.96 | 0.15 | 21.21 | 0.13 | 10.95504 | 41.71183 | 0.61 |
| 351 | 19.9 | 0.04 | 18.2 | 0.04 | 10.95546 | 41.18159 | 0.49 |
| 352 | 24.03 | 0.25 | 22.3 | 0.17 | 10.96394 | 41.70918 | 0.28 |
| 353 | 19.86 | 0.06 | 18 | 0.05 | 10.96992 | 41.73792 | 0.40 |
| 354 | 22.28 | 0.12 | 20.65 | 0.11 | 10.97108 | 41.73510 | 0.47 |
| 355 | 23.43 | 0.21 | 22.05 | 0.16 | 10.97116 | 41.72493 | 0.89 |
| 356 | 20 | 0.04 | 18.13 | 0.04 | 10.97144 | 41.74330 | 0.33 |
| 357 | 22.88 | 0.17 | 21.45 | 0.19 | 10.97157 | 41.74482 | 0.35 |
| 358 | 23.45 | 0.16 | 22.36 | 0.17 | 10.97563 | 41.72822 | 0.38 |
| 359 | 21.77 | 0.12 | 19.79 | 0.08 | 10.98045 | 41.73055 | 0.36 |
| 360 | 23.38 | 0.16 | 20.98 | 0.05 | 10.98214 | 41.74335 | 0.35 |
| 361 | 24.36 | 0.19 | 21.96 | 0.12 | 11.20145 | 41.52709 | 0.47 |
| 362 | 20.08 | 0.03 | 19.47 | 0.04 | 11.21268 | 41.50833 | 0.89 |
| 363 | 24.59 | 0.18 | 23.29 | 0.17 | 11.22412 | 41.53292 | 0.33 |
| 364 | 20.93 | 0.08 | 20.2 | 0.09 | 11.22783 | 41.53334 | 0.35 |
| 365 | 24.04 | 0.21 | 23.74 | 0.29 | 11.24400 | 41.55950 | 0.36 |

TABLE 2. GALAXY CANDIDATES WITH B, V PHOTOMETRY

| No. | V | V err | B | B err | B-V | RA | Dec | a arcsec |
|---|---|---|---|---|---|---|---|---|
| 365 | 24.08 | 0.27 | 25.06 | 0.33 | 0.98 | 10.08748 | 41.26338 | 0.23 |
| 366 | 19.31 | 0.03 | 20.72 | 0.05 | 1.40 | 10.09793 | 41.26547 | 0.00 |
| 367 | 21.93 | 0.06 | 23.31 | 0.14 | 1.38 | 10.09801 | 41.26438 | 0.00 |
| 368 | 25.47 | 0.27 | 25.76 | 0.40 | 0.29 | 10.10895 | 41.25651 | 0.17 |
| 369 | 20.78 | 0.05 | 21.90 | 0.08 | 1.12 | 10.10969 | 41.23380 | 0.23 |
| 370 | 22.89 | 0.11 | 23.42 | 0.15 | 0.53 | 10.11046 | 41.26419 | 0.17 |
| 371 | 25.25 | 0.23 | 26.31 | 0.51 | 1.05 | 10.11108 | 41.25463 | 0.00 |
| 372 | 22.23 | 0.06 | 23.44 | 0.14 | 1.22 | 10.11703 | 41.22063 | 0.17 |
| 373 | 22.18 | 0.07 | 23.25 | 0.13 | 1.07 | 10.11920 | 41.24162 | 0.00 |
| 374 | 23.53 | 0.14 | 25.58 | 0.40 | 2.05 | 10.12172 | 41.26733 | 0.00 |
| 375 | 21.82 | 0.18 | 22.99 | 0.17 | 1.17 | 10.12807 | 41.28544 | 0.17 |
| 376 | 23.01 | 0.10 | 24.96 | 0.30 | 1.95 | 10.12808 | 41.24554 | 0.00 |
| 377 | 22.42 | 0.09 | 23.36 | 0.15 | 0.93 | 10.13124 | 41.25818 | 0.00 |
| 378 | 21.08 | 0.07 | 22.17 | 0.09 | 1.09 | 10.13197 | 41.27652 | 0.59 |
| 379 | 20.83 | 0.06 | 21.85 | 0.08 | 1.03 | 10.13455 | 41.23253 | 1.10 |
| 380 | 21.59 | 0.07 | 22.88 | 0.12 | 1.29 | 10.13552 | 41.23755 | 0.50 |
| 381 | 23.60 | 0.13 | 24.61 | 0.25 | 1.01 | 10.13682 | 41.23067 | 0.28 |
| 382 | 22.54 | 0.09 | 24.37 | 0.24 | 1.83 | 10.13761 | 41.26707 | 0.55 |
| 383 | 24.19 | 0.17 | 24.95 | 0.28 | 0.77 | 10.13934 | 41.24491 | 0.35 |
| 384 | 24.16 | 0.18 | 25.32 | 0.35 | 1.17 | 10.14064 | 41.26888 | 0.54 |
| 385 | 24.05 | 0.16 | 24.94 | 0.29 | 0.89 | 10.14274 | 41.28557 | 0.21 |
| 386 | 22.90 | 0.11 | 24.08 | 0.20 | 1.18 | 10.14636 | 41.25893 | 0.21 |
| 387 | 23.89 | 0.13 | 24.77 | 0.26 | 0.89 | 10.14736 | 41.28109 | 0.71 |
| 388 | 23.44 | 0.12 | 24.50 | 0.23 | 1.06 | 10.14793 | 41.24043 | 0.21 |
| 389 | 21.44 | 0.06 | 22.80 | 0.11 | 1.36 | 10.14897 | 41.24136 | 0.40 |
| 390 | 22.70 | 0.10 | 23.60 | 0.15 | 0.90 | 10.14995 | 41.23875 | 0.71 |
| 391 | 23.27 | 0.12 | 25.05 | 0.33 | 1.78 | 10.15219 | 41.27464 | 0.33 |
| 392 | 24.66 | 0.16 | 25.77 | 0.41 | 1.11 | 10.15268 | 41.23524 | 0.46 |
| 393 | 23.23 | 0.09 | 25.05 | 0.31 | 1.82 | 10.15531 | 41.24890 | 0.33 |
| 394 | 22.48 | 0.07 | 23.79 | 0.17 | 1.32 | 10.15542 | 41.27092 | 0.46 |
| 395 | 23.13 | 0.10 | 23.78 | 0.17 | 0.65 | 10.16072 | 41.26174 | 0.00 |
| 396 | 22.23 | 0.09 | 23.38 | 0.15 | 1.15 | 10.16558 | 41.24778 | 0.00 |
| 397 | 24.25 | 0.17 | 25.58 | 0.37 | 1.34 | 10.16936 | 41.25050 | 1.44 |
| 398 | 22.41 | 0.09 | 23.46 | 0.15 | 1.05 | 10.17199 | 41.24591 | 0.35 |
| 399 | 21.17 | 0.07 | 22.07 | 0.08 | 0.89 | 10.29630 | 40.64710 | 0.57 |
| 400 | 21.08 | 0.08 | 21.64 | 0.08 | 0.56 | 10.29919 | 40.63044 | 0.23 |
| 401 | 22.67 | 0.10 | 23.08 | 0.12 | 0.41 | 10.30014 | 40.66572 | 0.42 |
| 402 | 20.68 | 0.05 | 21.16 | 0.06 | 0.48 | 10.30446 | 40.64761 | 0.22 |
| 403 | 23.53 | 0.30 | 23.79 | 0.34 | 0.26 | 10.31218 | 40.63349 | 0.51 |
| 404 | 21.76 | 0.07 | 21.94 | 0.08 | 0.18 | 10.31224 | 40.62340 | 0.40 |
| 405 | 21.08 | 0.07 | 21.58 | 0.07 | 0.50 | 10.33634 | 40.64476 | 0.25 |
| 406 | 21.32 | 0.07 | 22.43 | 0.10 | 1.11 | 10.34090 | 40.65846 | 0.62 |
| 407 | 21.56 | 0.08 | 23.14 | 0.15 | 1.58 | 10.43011 | 40.82936 | 0.00 |
| 408 | 22.57 | 0.08 | 24.11 | 0.21 | 1.54 | 10.46514 | 40.84981 | 0.00 |
| 409 | 22.72 | 0.07 | 23.62 | 0.16 | 0.90 | 10.47787 | 40.87130 | 1.44 |
| 410 | 21.26 | 0.06 | 22.77 | 0.11 | 1.51 | 10.97938 | 41.50411 | 0.39 |
| 411 | 23.00 | 0.10 | 24.16 | 0.19 | 1.15 | 11.08698 | 41.47984 | 1.24 |
| 412 | 20.77 | 0.05 | 21.92 | 0.08 | 1.15 | 11.09352 | 41.48563 | 0.39 |
| 413 | 21.96 | 0.09 | 22.51 | 0.10 | 0.55 | 11.10729 | 41.47571 | 0.25 |
| 414 | 20.94 | 0.06 | 22.02 | 0.08 | 1.09 | 11.10951 | 41.52378 | 0.31 |
| 415 | 19.67 | 0.04 | 21.22 | 0.07 | 1.56 | 11.10960 | 41.52484 | 0.35 |
| 416 | 23.33 | 0.14 | 25.68 | 0.46 | 2.35 | 11.11778 | 41.52934 | 0.44 |
| 417 | 21.72 | 0.06 | 23.11 | 0.14 | 1.38 | 11.15990 | 42.09064 | 0.40 |
| 418 | 20.60 | 0.03 | 21.65 | 0.07 | 1.05 | 11.16386 | 42.09337 | 0.25 |
| 419 | 25.22 | 0.32 | 26.09 | 0.47 | 0.88 | 11.16592 | 42.08580 | 0.62 |
| 420 | 21.80 | 0.06 | 23.09 | 0.14 | 1.30 | 11.16831 | 42.09802 | 1.24 |
| 421 | 21.95 | 0.10 | 23.14 | 0.17 | 1.19 | 11.17229 | 42.06877 | 0.39 |
| 422 | 23.52 | 0.22 | 25.00 | 0.48 | 1.48 | 11.17230 | 42.08210 | 0.25 |
| 423 | 22.23 | 0.09 | 23.72 | 0.20 | 1.49 | 11.17241 | 42.07090 | 0.31 |
| 424 | 22.75 | 0.13 | 23.81 | 0.20 | 1.06 | 11.17438 | 42.07658 | 0.35 |
| 425 | 21.03 | 0.06 | 21.34 | 0.07 | 0.31 | 11.17456 | 41.49476 | 0.40 |
| 426 | 21.00 | 0.05 | 22.31 | 0.09 | 1.32 | 11.17619 | 42.07395 | 0.00 |
| 427 | 22.60 | 0.08 | 23.71 | 0.16 | 1.12 | 11.17736 | 42.06692 | 0.00 |
| 428 | 23.16 | 0.10 | 24.27 | 0.22 | 1.11 | 11.17750 | 42.07386 | 0.15 |
| 429 | 20.60 | 0.04 | 21.44 | 0.06 | 0.84 | 11.17861 | 42.09037 | 0.09 |
| 430 | 22.53 | 0.06 | 23.44 | 0.14 | 0.91 | 11.18106 | 42.10286 | 0.11 |
| 431 | 22.72 | 0.09 | 23.96 | 0.19 | 1.24 | 11.18122 | 41.46541 | 0.31 |
| 432 | 23.57 | 0.15 | 24.72 | 0.27 | 1.15 | 11.18566 | 42.05368 | 0.18 |
| 433 | 20.17 | 0.05 | 21.44 | 0.10 | 1.27 | 11.18699 | 41.47802 | 0.44 |
| 434 | 18.41 | 0.02 | 19.14 | 0.04 | 0.73 | 11.18733 | 41.46392 | 0.00 |
| 435 | 20.67 | 0.05 | 21.19 | 0.06 | 0.52 | 11.19729 | 41.46833 | 0.00 |
| 436 | 20.99 | 0.04 | 22.53 | 0.10 | 1.54 | 11.20045 | 42.08489 | 0.20 |
| 437 | 20.62 | 0.07 | 21.11 | 0.06 | 0.50 | 11.20338 | 41.45345 | 0.09 |
| 438 | 21.00 | 0.05 | 21.44 | 0.06 | 0.44 | 11.21286 | 42.09805 | 0.21 |
| 439 | 19.86 | 0.13 | 20.14 | 0.18 | 0.28 | 11.21331 | 41.49002 | 0.21 |
| 440 | 22.11 | 0.22 | 22.81 | 0.18 | 0.71 | 11.21629 | 42.11021 | 0.43 |
| 441 | 20.74 | 0.06 | 21.56 | 0.08 | 0.81 | 11.22120 | 42.10682 | 0.00 |
| 442 | 20.99 | 0.13 | 21.95 | 0.17 | 0.96 | 11.22975 | 42.09623 | 0.00 |
| 443 | 21.94 | 0.12 | 22.87 | 0.15 | 0.93 | 11.23333 | 42.08410 | 0.00 |
| 444 | 22.50 | 0.10 | 23.61 | 0.15 | 1.11 | 11.25088 | 41.64896 | 0.43 |
| 445 | 22.41 | 0.10 | 22.77 | 0.10 | 0.36 | 11.27479 | 41.63987 | 0.00 |
| 446 | 21.87 | 0.09 | 23.81 | 0.19 | 1.94 | 11.29061 | 41.62598 | 0.00 |
| 447 | 23.72 | 0.20 | 24.28 | 0.23 | 0.56 | 11.29356 | 41.62171 | 0.30 |
| 448 | 21.42 | 0.08 | 22.16 | 0.09 | 0.74 | 11.29637 | 41.61253 | 0.15 |
| 449 | 20.21 | 0.05 | 20.66 | 0.05 | 0.46 | 11.30843 | 41.63929 | 0.30 |